\documentclass[showpacs,preprint]{revtex4}

\usepackage{amsmath}
\usepackage{amssymb}
\usepackage{graphicx}
\usepackage{psfrag}

\newcommand{\comm}[2]{\left[#1,#2\right]}

\newcommand{\vac}{\left|\,0\,\right\rangle}
\newcommand{\ket}[1]{\left|#1\right\rangle}

\newcommand{\bs}[1]{\boldsymbol{#1}}

\allowdisplaybreaks[4]

\def\ie{{\em i.e.},\ }

\def\ea{{\em et al.}}
\def\b{{\text{b}}}
\def\r{{\text{r}}}
\def\g{{\text{g}}}

\def\y{{\text{y}}}

\begin{document}

\title{Complementary colors of colorons: 
the elementary excitations of the SU(3) Haldane--Shastry model}

\author{Dirk Schuricht and Martin Greiter}

\affiliation{Institut f\"ur Theorie der Kondensierten Materie,
  Universit\"at Karlsruhe, Postfach 6980, 76128 Karlsruhe, Germany}

\pagestyle{plain}

\begin{abstract}
  We propose two possible trial wave functions for the elementary
  excitations of the SU(3) Haldane--Shastry model, but then argue on
  very general grounds that only one or the other can be a valid
  excitation.  We then prove explicitly that the trial wave function
  describing a coloron excitation which transforms according to
  representation $\bar{3}$ under SU(3) rotations if the spins of the
  original model transform according to representation 3, is exact.
  If a basis for the spins on the chain is spanned by the colors blue,
  red, and green, a basis for the coloron excitations is hence given
  by the complementary colors yellow, cyan, and magenta.  We derive
  the dispersion and show that colorons exist only in one third of the
  Brillouin zone. We further obtain the exclusion statistics among
  polarized colorons and discuss the generalization to SU($n$).
\end{abstract}



\maketitle

In recent years, there has been a substantial interest in models in
condensed matter physics with symmetry groups larger than SU(2), the
group underlying the usual spin algebra.  In particular, models with
SU(4) symmetry have been subject to detailed
investigations~\cite{LiMaShiZhang98,FrischmuthMilaTroyer99,AzariaGogolinLecheminantNerseyan99,LiMaShiZhang99,ItoiQinAffleck00,GuLi02},
as such models can be realized experimentally in transition-metal
oxides~\cite{IsobeUeda96,FujiiNakaoYosihamaNishiNakajimaKakuraiIsobeUedaSawa97,TokuraNagaosa00}
where the electron spin is coupled to orbital degrees of freedom.
Honerkamp and Hofstetter~\cite{HonerkampHofstetter04} have
investigated the SU($n$) generalization of the Hubbard model,
proposing possible experimental realizations in ultracold fermion
systems~\cite{AbrahamMcAlexanderGertonHuletCoteDalgarno97,RegalJin03}.

\begin{figure}[t]
\begin{center}
\includegraphics[scale=0.16]{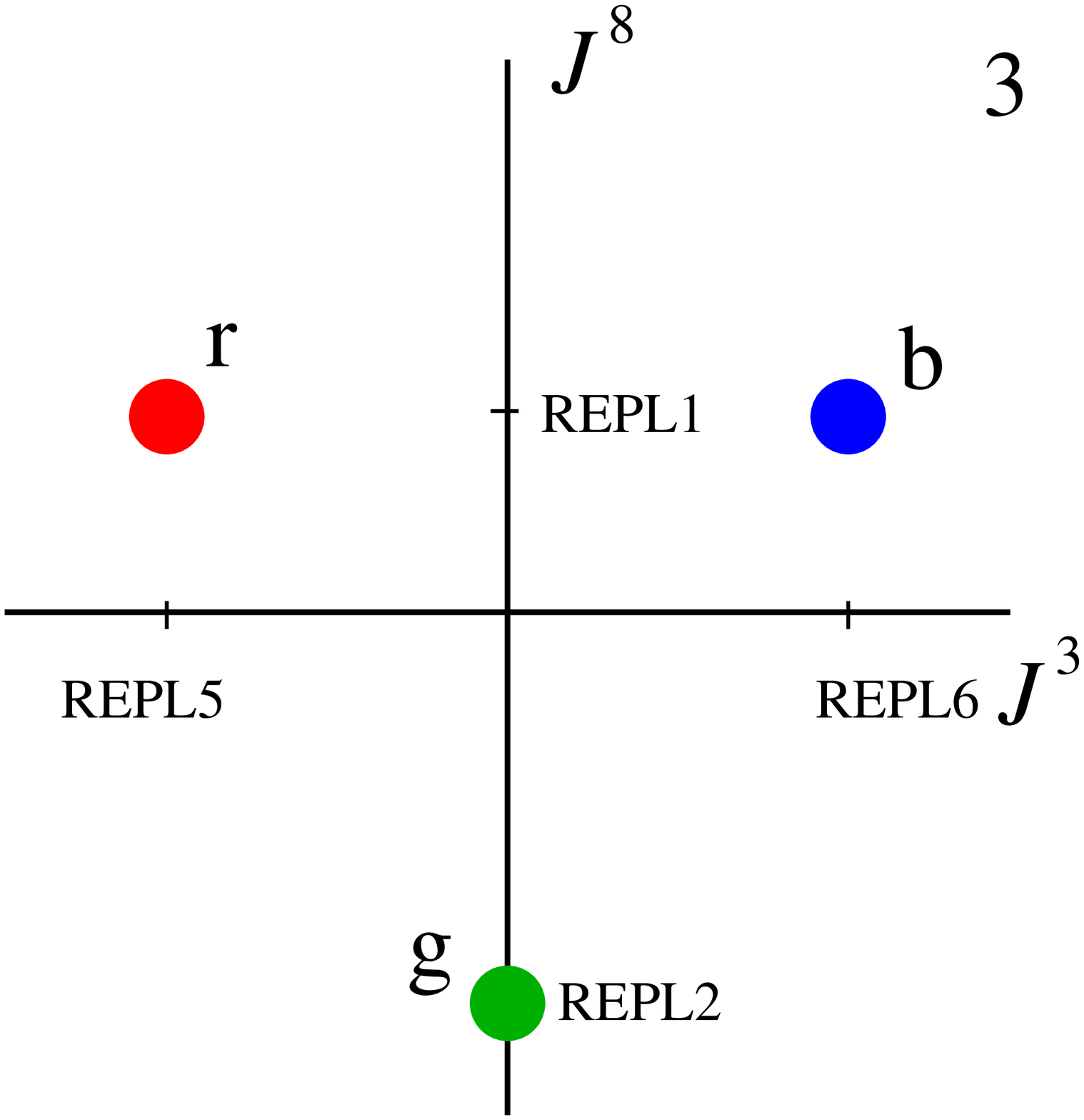}\hspace{2mm}
\includegraphics[scale=0.16]{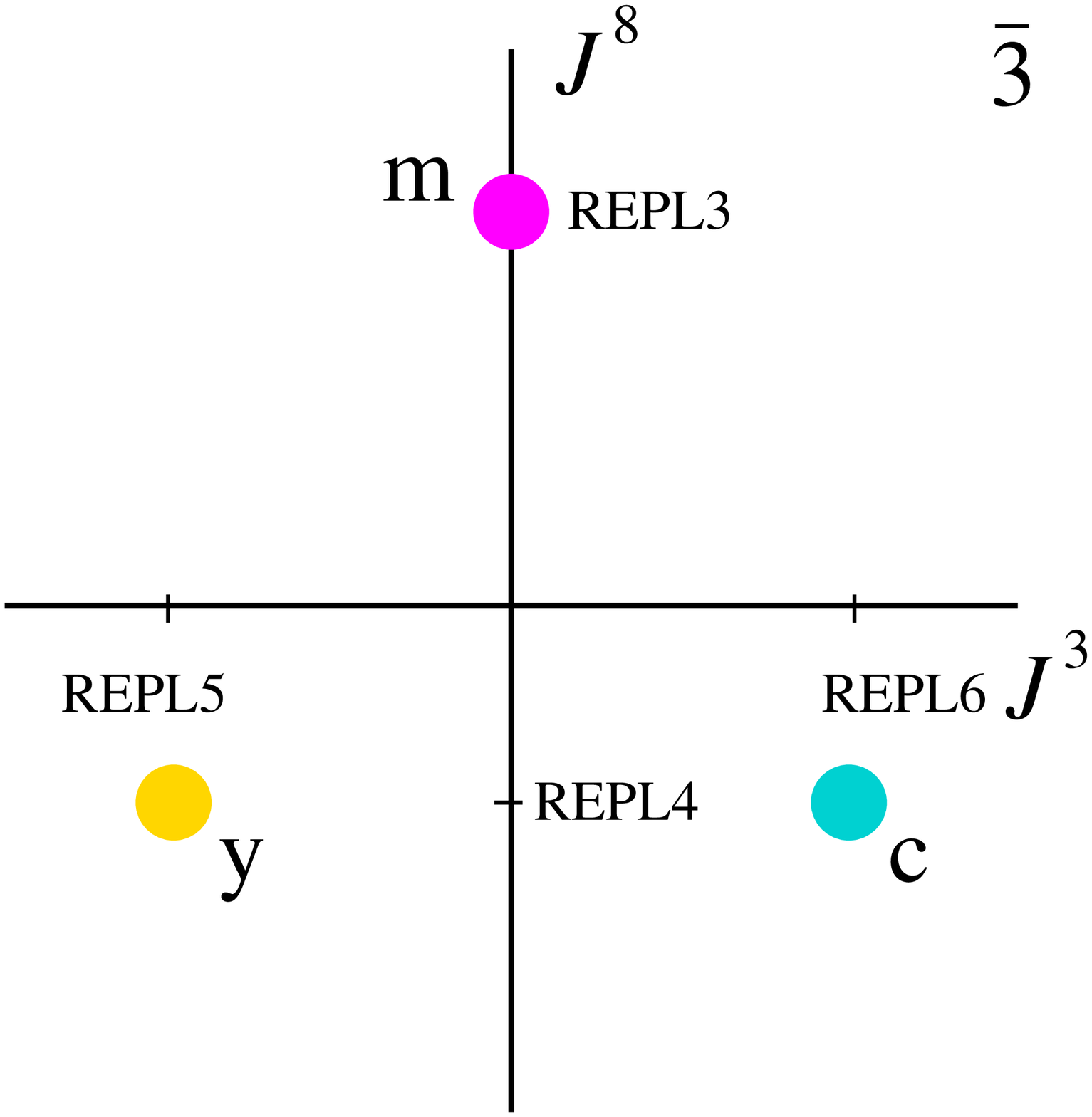}
\end{center}
\caption{Weight diagrams of the SU(3) representations $3$ and
$\bar{3}$. $J^3$ and $J^8$ denote the diagonal generators~\cite{Georgi82}.} 
\label{fig:reps}
\end{figure}

In this Letter, we will investigate yet another model in this general
class, the SU(3) Haldane--Shastry model
(HSM)~\cite{Haldane88,Shastry88,Haldane91prl1,HaldaneHaTalstraBernardPasquier92,Kawakami92prb1,Kawakami92prb2,HaHaldane92,HaHaldane93},
which serves as a paradigm for not only the SU(3) spin chain, but, as
we shall see, also illustrates some very general properties of SU($n$)
chains.  In particular, we derive the quantum numbers of the
elementary, fractionally quantized excitations, the analogs of the
spinon excitations in SU(2), which we call colorons for SU(3).  One of
the key results is that these excitations transform under the SU(3)
representation {\it conjugate} to the representation of the original
SU(3) spins localized at the sites of the chain (see
Fig.~\ref{fig:reps}).  First, this result reveals a new feature of
fractional quantization, namely the occurence of the conjugate
representation or complementary colors.  This feature is meaningless
in the case of SU(2) spin chains or the fractional quantum Hall
effect, but significant in all other instances of fractional
quantization in SU($n$) chains and possible higher dimensional
liquids, regardless of model specifics. Second, as a consequence of
their complementary colors the coloron excitations exist only in one
third of the Brillouin zone (BZ). This is reflected in the dynamical
structure factor (DSF) calculated exactly by Yamamoto
\ea~\cite{YamamotoSaigaArikawaKuramoto00prl,YamamotoSaigaArikawaKuramoto00jpsj}.
We hence provide an explicit interpretation of the DSF in terms of
wave functions, which can be used as starting point for the
investigation of other SU(3) spin chains as well.  In our analysis,
however, we will focus exclusively on the SU(3) HSM, as this is the
simplest model in which this general result can be illustrated through
exact solution, and thereby demonstrate the amenability of the SU(3)
HSM to exact solution in terms of explicit wave functions.

\begin{figure}[b]
\begin{center}
\includegraphics[scale=0.16]{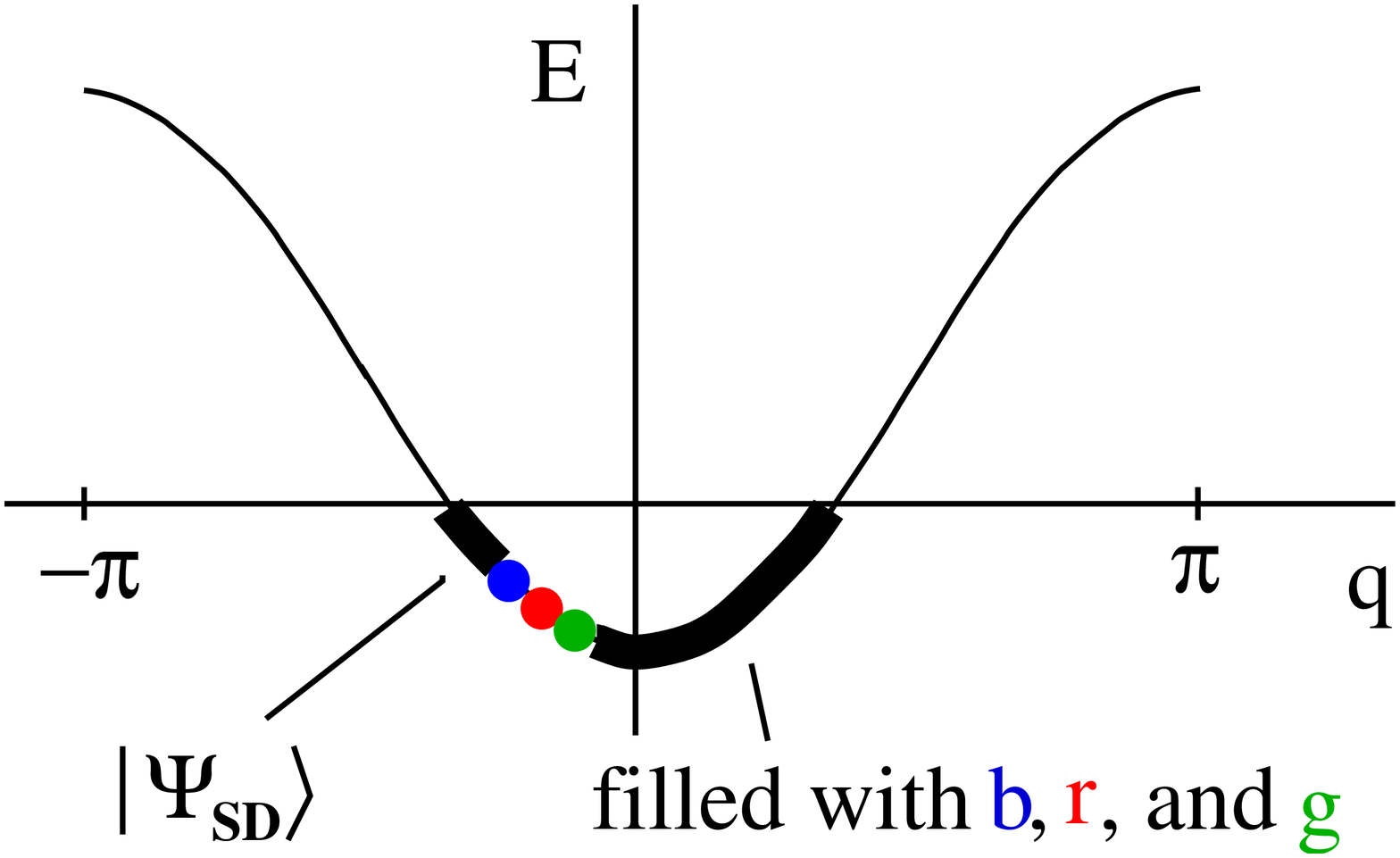}\hspace{6mm}
\includegraphics[scale=0.13]{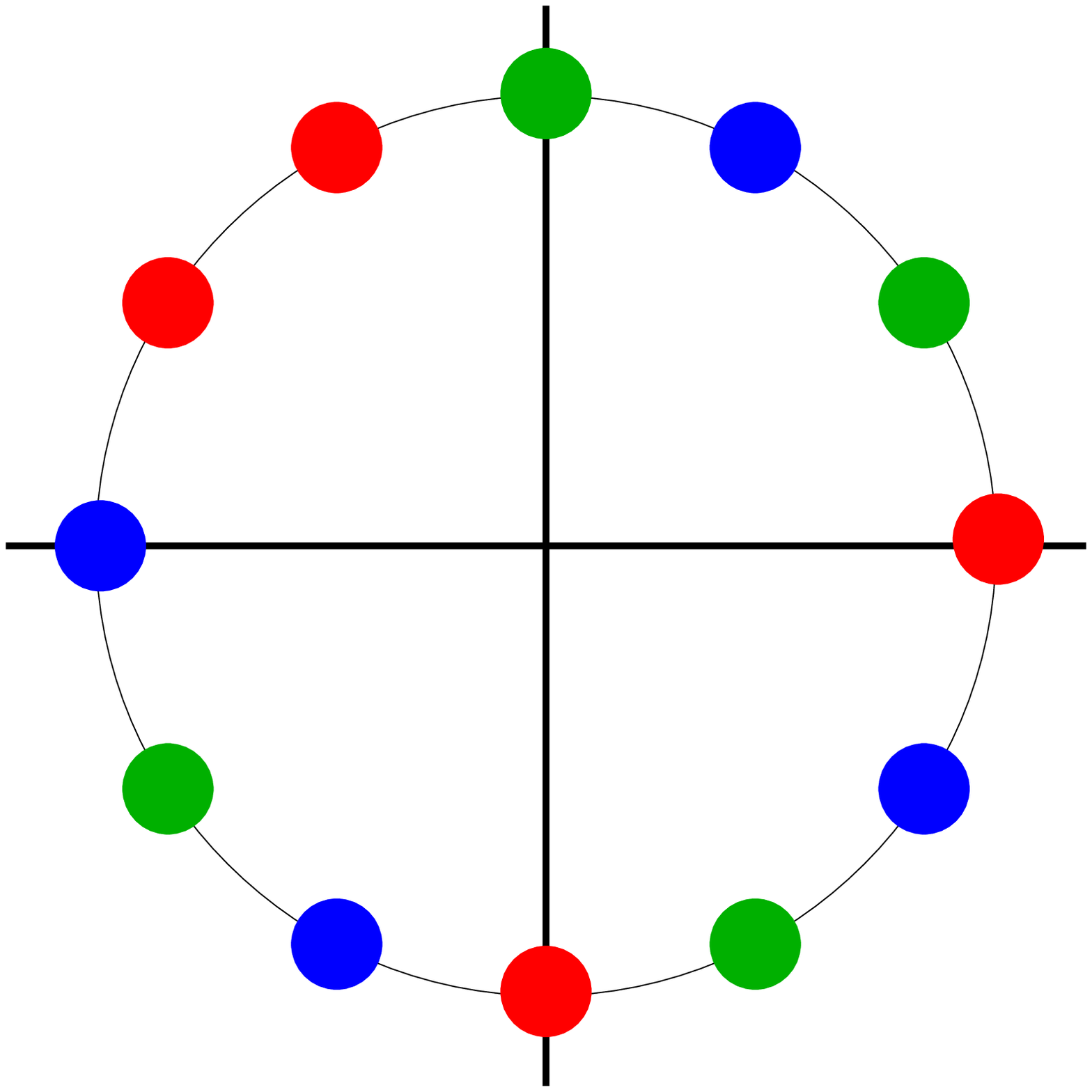}
\end{center}
\caption{a) Total antisymmetric $N$-particle state. 
b) Typical configuration in $\ket{\Psi_0}$.}
\label{fig:sd}
\end{figure}

The SU(3) $1/r^2$ or Haldane--Shastry
model~\cite{Kawakami92prb1,HaHaldane92} is most conveniently
formulated by embedding the one-dimensional chain with periodic
boundary conditions into the complex plane by mapping it onto the unit
circle with the SU(3) spins located at complex positions
$\eta_\alpha=\exp\!\left(i\frac{2\pi}{N}\alpha\right)$, where $N$
denotes the number of sites and $\alpha=1,\ldots,N$.  The Hamiltonian
is given by
\begin{equation}
  \label{eq:su3ham}
  H_{\text{SU(3)}}
  =\left(\frac{2\pi}{N}\right)^{\!\!2}
  \sum^N_{\alpha<\beta}\frac{\bs{J}_{\alpha}\!\cdot\!
    \bs{J}_{\beta}}{\vert \eta_{\alpha}-\eta_{\beta}\vert^2},
\end{equation}
where $\bs{J}_{\alpha}=\frac{1}{2}\sum_{\sigma\tau}
c_{\alpha\sigma}^{\dagger} \bs{\lambda}_{\sigma\tau}
c_{\alpha\tau}^{\phantom{\dagger}}$ is the 8-dimensional SU(3) spin
vector, $\bs{\lambda}$ a vector consisting of the 8 Gell-Mann
matrices~\cite{Georgi82}, and $\sigma$ and $\tau$ are SU(3) spin or
color indices, which take the values blue (b), red (r), or green (g)
(see Fig.~\ref{fig:reps}).  For all practical purposes, it is
convenient to express $H_{\text{SU(3)}}$ directly in terms of
colorflip operators $e_{\alpha}^{\sigma\tau}\equiv
c_{\alpha\sigma}^{\dagger} c_{\alpha\tau}^{\phantom{\dagger}}$:
\begin{equation}
  H_{\text{SU(3)}}
  =\frac{2\pi^2}{N^2}  
  \sum^N_{\alpha<\beta}\sum_{\sigma\tau}\,
  \frac{\:\!e_\alpha^{\sigma\tau}\,e_\beta^{\tau\sigma}-\frac{1}{27}\:\!}
  {\vert \eta_{\alpha}-\eta_{\beta}\vert^2},
  \label{eq:su3hamiltonian}
\end{equation}
where the color sum includes terms with $\sigma=\tau$. The ground
state of $H_{\text{SU(3)}}$ for $N=3M$ ($M$ integer) is most easily
formulated by Gutzwiller projection of a filled band (or Slater
determinant (SD) state) containing a total of $N$ SU(3) particles
obeying Fermi statistics (see Fig.~\ref{fig:sd}a)
\begin{equation}
  \ket{\Psi_0}=P_ {\text{G}}\prod_{|q|\le q_{\text{F}}}
  c_{q\text{b}}^\dagger\,c_{q\text{r}}^\dagger\,c_{q\text{g}}^\dagger
  \ket{0}
  \equiv P_{\text{G}}\ket{\Psi_{\text{SD}}^N}\!.
  \label{eq:su3-nnhgroundstate}
\end{equation}
The Gutzwiller projector $P_{\text{G}}$ eliminates configurations with
more than one particle on any site, and, as the total number of
particles equals the total number of sites, thereby effectively
enforces single occupancy on all sites.  As $\ket{\Psi_{\text{SD}}^N}$
is an SU(3) singlet by construction and $P_{\text{G}}$ commutes with
SU(3) rotations, $\ket{\Psi_0}$ is an SU(3) singlet as well.  The
model is fully integrable even for a finite number of sites; the
algebra of the (infinite) number of conserved quantities is generated
by the SU(3) total spin and rapidity operators
\begin{equation}
  \label{eq:jtotlambda}
  \bs{J}=\sum_{\alpha=1}^N \bs{J}_\alpha,\ \
  {\Lambda}^a=\frac{1}{2}\sum_{\alpha\neq\beta}^N\,
  \frac{\eta_\alpha + \eta_\beta}{\eta_\alpha - \eta_\beta}\,
  f^{abc}{J}_\alpha^b {J}_\beta^c,
\end{equation}
which commute with the Hamiltonian but do not commute mutually
($a=1,\ldots,8$ and $f^{abc}$ are the structure constants of SU(3)
defined through $\comm{\lambda^a}{\lambda^b}=2f^{abc}\lambda^c$).

If one interprets the state $\ket{0_\g}\equiv\prod_{\alpha=1}^N
c_{\alpha\g}\vac$ as a reference state and the colorflip operators
$e^{\b\g}$ and $e^{\r\g}$ as ``particle creation operators'', the
ground state (\ref{eq:su3-nnhgroundstate}) can be rewritten
as~\cite{Kawakami92prb2,HaHaldane92}
\begin{equation}
  \ket{\Psi_0}=\sum_{\{z_i,w_k\}}\Psi_0[z_i;w_k]\;
  e_{z_1}^{\b\g}\ldots e_{z_M}^{\b\g}
  e_{w_1}^{\r\g}\ldots e_{w_M}^{\r\g}\ket{0_\g}\!,
\end{equation}
where the sum extends over all possible ways to distribute the
positions of the blue particles $z_1,\ldots,z_M$ and red particles
$w_1,\ldots,w_M$ over the $N$ sites.  The ground state wave function
is given by (see Fig.~\ref{fig:sd}b)
\begin{equation}
    \Psi_0[z_i;w_k]\equiv
    \prod^{M_1}_{i<j}(z_i-z_j)^2\prod^{M_2}_{k<l}(w_k-w_l)^2
    \prod_{i=1}^{M_1}\prod_{k=1}^{M_2}(z_i-w_k)
    \prod_{i=1}^{M_1}z_i\prod_{k=1}^{M_2}w_k
  \label{eq:su3-definitionpsi0}
\end{equation}
with $M_1=M_2=M$. The ground state energy is
\begin{equation}
  E_0=-\frac{\pi^2}{18}\left(N+\frac{7}{N}\right)\!.
  \label{eq:su3-gsenergyM=N/3}
\end{equation}
The total momentum, as defined through $e^{ip}=\Psi_0[\eta_1
z_i,\eta_1 w_k]/\Psi_0[z_i,w_k]$ with $\eta_1=\exp(i\frac{2\pi}{N})$,
is $p=0$ regardless of $M$ (not true for SU(2)).

\psfrag{REPL1}{$\scriptstyle \frac{1}{2\sqrt{3}}$}
\psfrag{REPL2}{$\scriptstyle \frac{-1}{\sqrt{3}}$}
\psfrag{REPL3}{$\scriptstyle \frac{1}{\sqrt{3}}$}
\psfrag{REPL4}{$\scriptstyle \frac{-1}{2\sqrt{3}}$}
\psfrag{REPL5}{$\scriptstyle -\frac{1}{2}$}
\psfrag{REPL6}{$\scriptstyle \frac{1}{2}$}

We now turn to the heart of our analysis, the elementary and
fractionally quantized excitations, which we call colorons.  In
principle, there are two possible, non-equivalent constructions for
localized excitations starting from (\ref{eq:su3-nnhgroundstate}).  We
may either create a particle with color $\sigma$ on a chain with
$N=3M+1$ sites before Gutzwiller projection,
\begin{equation}
  \ket{\Psi_{\gamma\sigma}^{\text{c}}}=
  P_{\text{G}}\;\!c_{\gamma\sigma}^\dagger\!\ket{\Psi_{\text{SD}}^{N-1}}\!,
  \label{eq:loctwocolorons}
\end{equation}
or annihilate a particle with color $\sigma$ on a chain with $N=3M-1$:
\begin{equation}
  \ket{\Psi_{\gamma\bar\sigma}^\text{a}}=P_{\text{G}}\;\!
  c_{\gamma\sigma}^{\phantom{\dagger}}\!\ket{\Psi_{\text{SD}}^{N+1}}\!.
  \label{eq:loccoloron}
\end{equation}
In both cases, $c_{\gamma\sigma}^\dagger$ or
$c_{\gamma\sigma}^{\phantom{\dagger}}$ creates an inhomogeneity in
color and charge before projection.  The projection once again
enforces one particle per site and thereby removes the charge
inhomogeneity, while it commutes with color and thereby preserves the
color inhomogeneity.  Consequently, the trial states
$\ket{\Psi_{\gamma\sigma}^{\text{c}}}$ and
$\ket{\Psi_{\gamma\bar\sigma}^\text{a}}$ describe localized
``excitations'' of color $\sigma$ or complementary color $\bar\sigma$,
respectively, at lattice site $\eta_\gamma$.  Of course, since
$H_{\text{SU(3)}}$ is translationally invariant, we expect neither of
them, but only momentum eigenstates constructed from them via
\begin{equation}
  \ket{\Psi_n}
  \equiv\sum_{\gamma=1}^N e^{-i\frac{2\pi}{N}\gamma n}\ket{\Psi_\gamma}
  =\sum_{\gamma=1}^N (\bar\eta_\gamma)^n \ket{\Psi_\gamma},
  \label{eq:ncoloron}
\end{equation}
where $n$ is a momentum quantum number, to be eigenstates.

The important thing to realize now is that only 
(\ref{eq:loctwocolorons}) {or} (\ref{eq:loccoloron}), but not both,
can describe a valid excitation.  To see this, let us assume
$\sigma=\b$ for ease in presentation, and note that
(\ref{eq:loctwocolorons}) is apart from a normalization factor
equivalent to
\begin{equation}
  \ket{\Psi_{\gamma\b}^{\text{c}}}=
   P_{\text{G}}\;\!c_{\gamma\r}^{\phantom{\dagger}}
  c_{\gamma\g}^{\phantom{\dagger}}\ket{\Psi_{\text{SD}}^{N+2}},
  \label{eq:loctwocolorons2}
\end{equation}
\ie creation of a blue particle before projection is tantamount to
annihilation of both a red and a green particle at site $\eta_\gamma$.
If momentum eigenstates constructed from
$\ket{\Psi_{\gamma\bar\sigma}^\text{a}}$ via (\ref{eq:ncoloron}) were
energy eigenstates, the anti-red (cyan) and anti-green (magenta)
coloron excitations in (\ref{eq:loctwocolorons2}) would individually
seek to be momentum eigenstates, which implies that a trial wave
function forcing them to sit on the same site would not 
yield an energy eigenstate.  The same argument can be made the other
way round.

\begin{figure}[t]
 \begin{center}
 \setlength{\unitlength}{10pt}
 \begin{picture}(16,7)(-1,0)
 \put(0,1){\line(1,0){13.5}}
 \put(0,1){\line(0,1){5}}
 \qbezier[2000](4,1)(6,8)(8,1)
 \put(4,1){\makebox(0,0)[t]{\rule{0.3pt}{4pt}}}
 \put(8,1){\makebox(0,0)[t]{\rule{0.3pt}{4pt}}}
 \put(12,1){\makebox(0,0)[t]{\rule{0.3pt}{4pt}}}
 \put(0,0){\makebox(0,0){\small $0$}}
 \put(4,0){\makebox(0,0){\small $\frac{2\pi}{3}$}}
 \put(8,0){\makebox(0,0){\small $\frac{4\pi}{3}$}}
 \put(12,0){\makebox(0,0){\small $2\pi$}}
 \put(13.7,0.4){\makebox(0,0){\small $p$}}
 \put(-1.3,6){\makebox(0,0){\small $E(p)$}}
 \put(6,1){\makebox(0,0){\rule{40pt}{3pt}}}
 \put(6,1.5){\line(1,1){3}}
 \put(12,5.3){\makebox(0,0){\small allowed momenta}}
 \end{picture}
 \end{center}
 \caption{One-coloron dispersion relation.}
 \label{fig:dis}
\end{figure}
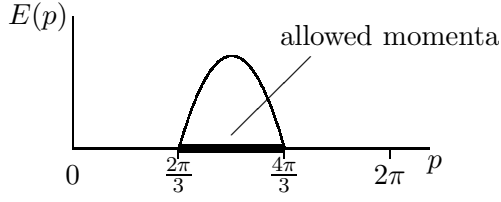

In the following, we show by explicit calculation that momentum
eigenstates constructed from $\ket{\Psi_{\gamma\bar\sigma}^\text{a}}$
via (\ref{eq:ncoloron}) are exact eigenstates of $H_{\text{SU(3)}}$
with momentum
\begin{equation}
  p=\frac{4}{3}\pi -\frac{2\pi}{N}\left(n+\frac{1}{3}\right)\!,
  \quad 0\le n\le \frac{N-2}{3},
\label{eq:coloronmomentum}
\end{equation}
(where $n$ is shifted with respect to $n$ in (\ref{eq:ncoloron}) by a
constant depending on which momenta are occupied in the Slater
determinant state) and energy
\begin{equation}
  E(p)=E_0+\frac{2}{9}\frac{\pi^2}{N^2} 
  +\frac{3}{4}\left(\frac{\pi^2}{9}-(p-\pi)^2 \right)\!
  \label{eq:colorondisperion}
\end{equation} 
(see Fig.~\ref{fig:dis}).  For $(N-2)/3<n<N$,
$\ket{\Psi_{n\bar\sigma}^\text{a}}$ vanishes identically.  The color
$\bar\sigma$ of the excitation, constructed by annihilation of a
particle of color $\sigma$ from an overall color singlet
$\ket{\Psi_{\text{SD}}^{N+1}}$, hence transforms according to the
representation $\bar 3$ conjugate to the fundamental representation
$3$ of the original particles on the sites of the chain.  This is
consistent with results on the spectrum of the SU(3)$_1$
Wess--Zumino--Witten (WZW) model obtained by Bouwknegt and
Schoutens~\cite{BouwknegtSchoutens96,Schoutens97}.  The momenta of the
coloron excitations $\ket{\Psi_{n\bar\sigma}}$ fill only one third of
the BZ, which is fully consistent with the calculations of the
DSF~\cite{YamamotoSaigaArikawaKuramoto00prl,YamamotoSaigaArikawaKuramoto00jpsj}.

For simplicity, we choose $\bar\sigma=\bar\b=\y$ (an anti-blue or
yellow coloron) and express
$\bigl|\Psi_{\gamma\bar\b}^\text{a}\bigr\rangle$ through the
corresponding wave function
\begin{equation}
  \Psi_\gamma[z_i;w_k]=
  \prod_{i=1}^{M_1} (\eta_\gamma-z_i)\;\Psi_0[z_i;w_k]\equiv\psi_\gamma\Psi_0.
  \label{eq:localizedcoloronwf}
\end{equation}
with $\Psi_0$ given by (\ref{eq:su3-definitionpsi0}) and
$M_1=(N-2)/3$, $M_2=(N+1)/3$.  The momentum eigenstate
$\bigl|\Psi_{n\bar\b}^\text{a}\bigr\rangle$ is then given by
\begin{equation}
  \Psi_n[z_i;w_k]=\sum_{\gamma=1}^N(\bar\eta_{\gamma})^n\;\Psi_\gamma[z_i;w_k],
  \label{eq:wfmomentumcoloron}
\end{equation}
which vanishes identically unless $0\le n\le M_1$.  To evaluate the
action of $H_{\text{SU(3)}}$ on (\ref{eq:wfmomentumcoloron}) we first
replace $e_\alpha^{\g\g}e_\beta^{\g\g}$ in (\ref{eq:su3hamiltonian})
by
$(1-e_\alpha^{\b\b}-e_\alpha^{\r\r})(1-e_\beta^{\b\b}-e_\beta^{\r\r})$
and treat each term separately~\cite{manuscriptinpreparationSG}.  For
example, the term
$[e_\alpha^{\b\g}e_\beta^{\g\b}\Psi_\gamma][z_i;w_k]$, which vanishes
unless one of the $z_i$'s is equal to $\eta_\alpha$, yields through
Taylor expansion
\begin{eqnarray}
\left[\sum_{\alpha\neq\beta}^N
\frac{e_\alpha^{\b\g}e_\beta^{\g\b}}{\vert\eta_\alpha-\eta_\beta\vert^2}
\Psi_\gamma\right]\!\![z_i;w_k]&=&\sum_{i=1}^{M_1}\sum_{m=0}^{N-1}
\frac{A_mz_i^{m+1}}{m!}
\frac{\partial^m}{\partial z_i^m}\frac{\Psi_\gamma}{z_i}\nonumber\\ 
&=&-\frac{N^3+12 N^2-123N+190}{324}\Psi_\gamma\nonumber\\
& &-\frac{N-3}{2}\sum_{i=1}^{M_1}\sum_{k=1}^{M_2}
\frac{z_i}{z_i-w_k}\Psi_\gamma
+\sum_{i\neq j}^{M_1}\frac{z_i^2}{(z_i-z_j)^2}\Psi_\gamma\nonumber\\ 
& &+2\sum_{i\neq j}^{M_1}\sum_{k=1}^{M_2}
\frac{z_i^2}{(z_i-z_j)(z_i-w_k)}\Psi_\gamma\nonumber\\
& &+\frac{1}{2}\sum_{i=1}^{M_1}\sum_{k\neq l}^{M_2}
\frac{z_i^2}{(z_i-w_k)(z_i-w_l)}\Psi_\gamma\label{eq:1331}\\ 
& &+\Psi_0\sum_{i=1}^{M_1}\!\!\left(\!\frac{z_i^2}{2}\partial^2_i +
\sum_{j\neq i}^{M_1} \frac{2z_i^2}{z_i-z_j}
\partial_i-\frac{N-3}{2}z_i\partial_i\!
\right)\!\psi_\gamma\nonumber\\ 
& &+\Psi_0\sum_{i=1}^{M_1}\sum_{k=1}^{M_2}\frac{z_i^2}{z_i-w_k}
\partial_i\psi_\gamma\nonumber,
\end{eqnarray}
where we have used $\text{deg}_{z_i}\Psi_\gamma[z_i;w_k]=N-1$ and
defined $\partial_i\equiv\partial/\partial z_i$,
$A_m\equiv-\sum_{\alpha=1}^{N-1} \eta_\alpha^2 (\eta_\alpha
-1)^{m-2}$.  Evaluation of the latter yields $A_0=(N-1)(N-5)/12$,
$A_1=-(N-3)/2$, $A_2=1$, and $A_m=0$ for $2<m\le
N-1$~\cite{BernevigGiulianoLaughlin01prb}.  The action of $\sum
e_\alpha^{\b\g}e_\beta^{\g\b}/|\eta_\alpha-\eta_\beta|^2$ on $\Psi_n$
is obtained by Fourier transformation of (\ref{eq:1331}); the first
six terms on the RHS can be treated in analogy to the SU(3) HS ground
state~\cite{HaHaldane92} and the calculations in the SU(2) HS
model~\cite{Haldane88,BernevigGiulianoLaughlin01prb}.

To mention another term,
\begin{eqnarray}
\left[\sum_{\alpha\neq\beta}^N
\frac{e_\alpha^{\b\r}e_\beta^{\r\b}}{\vert\eta_\alpha-\eta_\beta\vert^2}
\Psi_n\right]\!\![z_i;w_k]\nonumber
&=&\!\!\frac{1}{N}\sum_{\gamma=1}^N(\bar{\eta}_\gamma)^{n}
\sum_{i=1}^{M_1}\sum_{k=1}^{M_2}\frac{z_iw_k}{(z_i-w_k)^2}
\left(1+\frac{z_i-w_k}{\eta_\gamma-z_i}\right)\nonumber\\
& &\cdot\prod^{M_1}_{j\neq i}\frac{z_j-w_k}{z_j-z_i}
\prod^{M_2}_{l\neq k}\frac{w_l-z_i}{w_l-w_k}
\Psi_\gamma[z_i;w_k],
\label{eq:1221term}
\end{eqnarray}
splits into two parts: one, which can again be treated as for the
ground state, and another, the sum of which with the last term in
(\ref{eq:1331}) can be simplified using
\begin{displaymath}
  \begin{split}
    \sum_{i=1}^{M_1}\sum_{k=1}^{M_2}\frac{z_i}{z_i-w_k}\cdot
    &\hspace{-3pt}\left[w_k\prod^{M_1}_{j\neq i}\frac{z_j-w_k}{z_j-z_i}
      \prod^{M_2}_{l\neq k}\frac{w_l-z_i}{w_l-w_k}-z_i\right]
    \sum^N_{\gamma=1}(\bar{\eta}_\gamma)^{n}
    \prod^{M_1}_{j\neq i}(\eta_\gamma-z_j)\\
    &=\!\!\left[-\frac{1}{2}n(n+1)+\frac{1}{2}M_1(M_1+1)\right]
    \sum^N_{\gamma=1}(\bar{\eta}_\gamma)^{n}\prod^{M_1}_{i=1}(\eta_\gamma-z_i),
  \end{split}
\end{displaymath}
which is valid for $M_1\le M_2$~\cite{manuscriptinpreparationSG}.  In
this manner, the action of every term in (\ref{eq:su3hamiltonian}) on
(\ref{eq:wfmomentumcoloron}) can be evaluated, yielding that
(\ref{eq:wfmomentumcoloron}) is indeed an eigenstate with energy
(\ref{eq:colorondisperion}) and momentum (\ref{eq:coloronmomentum}).

It is straightforward to read off the quantum or exclusion
statistics~\cite{Haldane91prl2} of color-polarized colorons.  Consider
a chain with $N=3M-1$ sites and a single yellow coloron.  According to
(\ref{eq:su3-nnhgroundstate}), (\ref{eq:loccoloron}), and
(\ref{eq:ncoloron}), there are as many single particle orbitals
available to the coloron as there are blue particles in the Slater
determinant state, that is, $M$.  If we now were to create three
additional yellow colorons, the Slater determinant state would have to
contain three more particles, one of each color.  This implies there
would be one additional orbital, while the three additional colorons
would occupy three orbitals, meaning that the number of orbitals
available for our original coloron would be reduced by two.  The
statistical parameter is hence given by $g=2/3$.  The fractional
statistics manifests itself further in the exponents of the algebraic
decay of the form factor of the
DSF~\cite{YamamotoSaigaArikawaKuramoto00prl,YamamotoSaigaArikawaKuramoto00jpsj},
in the thermodynamics of the
model~\cite{KuramotoKato95,KatoKuramoto96}, as well as the S-matrix
evaluated with the asymptotic Bethe Ansatz~\cite{Essler95}.  A similar
exclusion statistics exists in the conformal field theory spectrum of
WZW models~\cite{Schoutens97,BouwknegtSchoutens99}.  The general
exclusion statistics and state counting for SU($n$) ($n\ge 3$) spin
chains is highly non-trivial and will be subject of a future
publication~\cite{manuscriptinpreparationGS}.

The results derived here generalize directly to the SU($n$)
HSM~\cite{manuscriptinpreparationSG}.  The eigenstates describing a
single fractionally quantized SU($n$) spinon excitation are given by
the SU($n$) generalization of (\ref{eq:loccoloron}).  If the particles
on the sites of the chain transform under the fundamental
representation $n$ of SU($n$), the spinons transform under the
conjugate representation $\bar{n}$.  The allowed momenta of the
spinons are restricted to $p\in [\pi-\frac{\pi}{n},\pi+\frac{\pi}{n}]$
(or $[-\frac{\pi}{n},\frac{\pi}{n}]$ if $n$ even and $(N+1)/n$ odd),
and hence fill only the $n^\text{th}$ part of the 
BZ, with dispersion
\begin{equation}
E^n(p)=E_0^n+\frac{n^2-1}{12n}\frac{\pi^2}{N^2}+
\frac{n}{4}\left(\frac{\pi^2}{n^2}-(p-\pi)^2\right)\!,
\end{equation}
where $E_0^n$ is the ground state energy~\cite{Kawakami92prb2,HaHaldane92}.
The statistical parameter for polarized spinons is given by $(n-1)/n$.  

In conclusion, we have used the HSM to make the case that the
elementary excitations of SU($n$), but in particular SU(3), spin
chains transform under the representation conjugate to the
representation of the SU($n$) spins on the chain, exist only in one
$n^\text{th}$ of the Brillouin zone, and obey fractional statistics
with $g=(n-1)/n$.

One of us (DS) was supported by the German Research Foundation (DFG) through
GK 284.  The other (MG) would like to thank the organizers of the 2003
Amsterdam Summer Workshop on Flux, Charge, Topology and Statistics, where this
work was partially inspired.

\end{document}